\documentclass{WileyMSP-template}

\usepackage[normalem]{ulem}

\begin{document}

\pagestyle{fancy}
\rhead{\includegraphics[width=2.5cm]{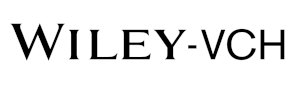}}

\title{Overcoming Computational Bottlenecks in Quantum Hydrodynamics: A Volume-Based Integral Formalism}

\maketitle


\author{Christos Mystilidis}
\author{Christos Tserkezis}
\author{Guy A. E. Vandenbosch}
\author{N. Asger Mortensen}
\author{Xuezhi Zheng*}


\dedication{}

\begin{affiliations}
Dr. Christos Mystilidis \\
Email address: christos\_mystilidis@mail.ntua.gr \\
Address: School of Electrical and Computer Engineering, National Technical University of Athens, 9 Iroon Polytechniou Street GR 157-73 Zografou, Athens, Greece \\

Prof.  Christos Tserkezis\\
Address: POLIMA---Center for Polariton-driven Light-–Matter
Interactions, University of Southern Denmark, DK-5230 Odense, Denmark.\\

Prof. Guy A. E. Vandenbosch\\
Address: WaveCoRE Division, Department of Electrical Engineering (ESAT), KU Leuven, Kasteelpark Arenberg 10, box 2444, Leuven, B-3001, Belgium.\\

Prof. N. Asger Mortensen\\
Address: POLIMA----Center for Polariton-driven Light-–Matter
Interactions, University of Southern Denmark, DK-5230 Odense, Denmark.
D-IAS: Danish Institute for Advanced Study, University of Southern Denmark, Campusvej 55, DK-5230 Odense M, Denmark\\

Prof. Xuezhi Zheng\\
Email Address: xzheng@nuaa.edu.cn \\
Address: College of Electronics and Information Engineering, Nanjing University of Aeronautics and Astronautics, Nanjing, China \\

\end{affiliations}



\begin{abstract}
Mesoscopic models of the optical response of metals have emerged as fundamental building blocks in quantum plasmonics, in principle overcoming the computational bottlenecks of ab initio techniques by implementing aspects of the atomistic description of the metal in otherwise classical calculations. Nonetheless, even these approaches are eventually hindered by demanding computations due to sophisticated material response. Here, this issue is addressed for the advanced Self-Consistent Hydrodynamic Drude Model (SC-HDM), which captures both nonlocal electron dynamics and electron spill-out, through a Volume Integral Equation (VIE) method. Adopting an IE-based method shifts perspective from the commonly employed Differential Equation (DE)-based ones, demonstrating significant computational efficiency. The VIE approach is a valuable methodological scaffold: It addresses SC-HDM and simpler models, but can also be adapted to more advanced ones. For spherical nanoparticles (NPs), using the inherent symmetries, similar performance for three increasingly complicated material models is achieved, breaking the taboo that increased sophistication in material response requires taxing simulations. Mesoscopic material-response functions can be readily extracted from the VIE implementation, thus circumventing the need for lengthy microscopic calculations. This method opens a new way of modeling quantum hydrodynamic NPs and will serve as essential benchmarking tool for recipes addressing more complicated geometries. 

\end{abstract}
 

\section{Introduction}
\label{sec:intro}

The effort to localize light at ever decreasing volumes far below the diffraction limit, and the exploitation of the concomitant large field enhancements have long monopolized developments in nanophotonics, have established metals or heavily-doped semiconductors as quintessential components for realizing field localization, and have defined the very research discipline of plasmonics~\cite{Gramotnev2010}. These phenomena, combined with the ease of engineering plasmonic devices~\cite{Prodan03}, invite fruitful synergies of plasmonics with a wide range of independent disciplines. Characteristic examples are the translation of RF antennas operating at the microwave regime to the optical and nanoscale one~\cite{Novotny11,Giannini11}, even leading to complete wireless systems~\cite{Merlo16}, the amplification of efficiency of photovoltaic cells~\cite{Atwater10,Jang16}, the design of ultrafast and compact nanolasers~\cite{Ma12,Azzam20}, and the plasmonic facilitation of chemical reactions~\cite{Cortes2022}. 

In order to achieve stronger localization and field enhancement, classical electrodynamics, which has been the main vehicle of plasmonics, dictates the monotonous shrinking of the dimensions of plasmonic devices and in particular of interparticle gaps. This prediction is the starting point~\cite{Zhu16} behind spectacular advancements in nanofabrication techniques that eventually led to the proliferation of experimental platforms for the observation of extreme optical phenomena~\cite{Scholl12,Savage12,Boroviks22}, such as the squeezing of light in sub-1 nm\textsuperscript{3} volumes~\cite{Benz16}. Such trend, however, sets serious doubts on the use of classical electrodynamics---which in the nanoplasmonics parlance is often called Local Response Approximation (LRA)---for the design and engineering of such platforms. When the dimensions of geometric characteristics are of the same scale as the wavelength of matter waves of electrons (sub-1 nm for typical metals~\cite{Mortensen21}), it is expected that quantum mechanical behavior progressively dominates the optical response. 

While the fully macroscopic approach becomes inaccurate, the fully microscopic approach, namely \emph{ab initio} techniques relying on the solution of the many-body Schr\"{o}dinger equation, or effective single-electron ones, such as the Time-Dependent Density Functional Theory (TD-DFT), are computationally prohibitive for all but the smallest systems of interest~\cite{Dawson22}. In this intermediate regime, where classical electrodynamics is interfacing with quantum mechanics~\cite{Mortensen21}, an alternative approach, where the material response is augmented by ad hoc integration of quantum mechanical effects may be preferable, sacrificing accuracy, but offering a computationally more appealing framework. 
This approach relies on the design of appropriate \emph{mesoscopic models}. Many such models have long been suggested, focusing on different quantum mechanical mechanisms. For example, the Quantum Corrected Model (QCM)~\cite{Esteban12,Esteban15} integrates quantum tunnelling by adding a conducting junction in the interparticle gap of metallic dimers. Kreibig's Size-Dependent Broadening (SDB) model~\cite{Kreibig74} offers a simple recipe to predict size-dependent damping in spherical NPs, which originates from surface-enhanced Landau damping and the reduction of the electron mean free path in the bulk. Recently, the Surface Response Model (SRM)~\cite{Feibelman82}, based on the mesoscopic material response functions known as \emph{Feibelman's $d$ parameters}, which can include nonlocality, electron spill-out, and Landau damping in a format familiar to classical electrodynamics, albeit with appreciably different boundary conditions, has gained traction~\cite{Christensen17,Yang19,RodriguezEcharri21,Babaze22,Eriksen24,Bundgaard24,Eremin25}.

Among the most popular realizations of mesoscopic electrodynamics one encounters models based on treating the interacting electron gas as a charged fluid, through a hydrodynamic description. In such approaches, which date back to the 1930s~\cite{Bloch33}, but only returned to relevance the last decade due to the emergence of quantum plasmonics as an independent research discipline, the well-known Drude model is augmented to include electron--electron interactions beyond collisional dynamics. In its most simple iteration, known as the Hydrodynamic Drude Model (HDM), it introduces the Pauli exclusion principle and eventually Thomas-Fermi screening in the optical response~\cite{Ciraci13b,Forstman86}. HDM has been tremendously successful in interpreting Electron Energy-Loss Spectroscopy (EELS) data of noble metals (such as gold and silver)~\cite{Raza13a} and the regularization of field singularities at the touching limit of metallic  nanoparticles (NPs)~\cite{Ciraci12}, a critical deficiency of classical electrodynamics~\cite{Romero06}. Nonetheless, it is the superposition and sometimes the competition of quantum effects that determines the full optical response from metallic systems~\cite{Christensen17}. In alkali metals (such as sodium--Na), the optical response is dominated by electron spill-out, and as such HDM is totally inaccurate, more so than even LRA, and has received fair criticism~\cite{Teperik13a}. The Self-Consistent Hydrodynamic Drude Model (SC-HDM; also known as Quantum Hydrodynamic Theory) was proposed as a solution to this problem~\cite{Toscano15}. Maintaining a framework familiar from DFT~\cite{Eguiluz75a,Eguiluz75b,Eguiluz76}, it allows for a progressive inclusion of additional sophistication in material response; the most conventional implementation~\cite{Toscano15,Ciraci16} contains gradient corrections to the kinetic energy of the electron gas and enriches its dynamics by adding approximations to exchange-correlation (XC) contributions. Note that additional sophistication may be added~\cite{Ciraci17,Baghramyan21,Zhou23}, establishing a hierarchy of progressively more complicated but also more accurate hydrodynamic models.

From the discussion above, a clear advantage of the SC-HDM ensues: \emph{modularity}. More than a model, it is a \emph{model generator}, where a hierarchy of models can be established depending on the details included in the material response. After the starting work~\cite{Toscano15}, significant theoretical effort has been carried out by Cirac\`{i} and coworkers~\cite{Ciraci16,Ciraci17,Baghramyan21} who were the first to describe the influence of the input ground-state density (which plays here a role equivalent to the conventional permittivity in classical electrodynamics) in the optical response and revealed stability issues that plague to an extent numerics in this model, building on the earlier work by Yan~\cite{Yan15}. Interrogations of this model in more complicated geometries than the archetypical sphere or cylinder ensued: core--shell nanostructures~\cite{Khalid18,Khalid19}; polygonal rods \cite{Ding17}; bow-tie nanoantennas~\cite{Ding18}; spherical and conical dimers \cite{Noor22,Zhang25}; considering both linear and nonlinear response~\cite{Khalid20,Noor22}, excited by plane waves but also quantum emitters~\cite{Ciraci19}. Yet, it is clear that the nanoplasmonics community has been far more cautious with adopting SC-HDM in comparison to its simpler sibling HDM, a surprising development given its---in principle---stronger predicting powers. We find two reasons behind such reservations. For one, SC-HDM (and more advanced models) introduce a \emph{withering out} of boundaries; due to spill-out, the plasmonic scatterer extends in the entire space, making analytical progress formidably tough. At the same time, it appears that SC-HDM experiences severe computational bottlenecks, requiring dedicated computing facilities~\cite{Toscano15}, or has not received sufficient ad hoc recipes. All the works referenced up until now use the commercial package COMSOL (albeit the works of Cirac\`{i} and coworkers utilize an optimized 2.5D approach~\cite{Ciraci13}, harvesting symmetries of axisymmetric structures and resolving the azimuth dependence of fields in terms of cylindrical harmonics). To the best of our knowledge, the only dedicated computational works on SC-HDM are the pioneering~\cite{Takeuci22,VidalCodina23}, the former a Finite-Difference Time-Domain (FDTD) method and the latter a hybridizable discontinuous Galerkin one, which have achieved significant progress since 2015, but in stark contrast to the rich computational literature in HDM~\cite{Hiremath12,Eremin18,Mystilidis23b,Du24}.

In this work, we develop a new theoretical method for the treatment of SC-HDM-described NPs. Leveraging the familiar from classical electrodynamics volume equivalence principle~\cite{Chew95}, we introduce an Integral Equation (IE)-based formalism. In this manner, we bypass a usual problem of multiphysics in general and semiclassical Differential Equation (DE)-based numerical methods in particular, which is the reconciliation of the vastly different characteristic length scales, a clear departure from previous theoretical efforts~\cite{Takeuci22,VidalCodina23}. As a first attempt, we apply the suggested formalism to the case of isolated spherical NPs. There, inspired by symmetry principles, we expand the working variable, namely the induced polarization density, in terms of vector spherical harmonics. This expansion is the key behind tremendous  relaxation of the computational demands by effectively rendering the problem one-dimensional (1D). With the presentation of this method, we argue that the bottlenecks that SC-HDM has experienced are \emph{artificial}. Very importantly, we reveal that SC-HDM can potentially display performance of the same order of execution time as fully classical electrodynamics. Further, we suggest a pipeline scheme, where SC-HDM can be used to extract Feibelman parameters (replacing effectively TD-DFT) with minimal computational cost and directly load them to SRM-related solvers. We go on to validate our approach step-by-step against an independent $\mathbf{S}$ matrix solver~\cite{Mystilidis23a} and unveil the rich physics of SC-HDM with respect to the size dependence of the electric-dipole plasmon and the emergence of the Bennett resonance (also known as multipole surface plasmon)---a fingerprint of systems where spill-out is admitted and arising precisely due to the dipole layer of the induced charge that is created across the interface in this case~\cite{Bennett70}.

\section{Theory}
The analysis below is given in the frequency domain, assuming and subsequently suppressing a $\mathrm{e}^{-i\omega t}$ time dependence. As per usual, $i=\sqrt{-1}$ is the imaginary unit, $\omega$ is the angular frequency, and $t$ is the time. Since we study natural materials at optical frequencies, we assume a permeability $\mu=\mu_0$, the vacuum magnetic permeability, throughout. 

\subsection{Quantum Corrected Material Response: The Self-Consistent Hydrodynamic Drude Model}
\label{sec:schdm}

Formally, the SC-HDM is defined as the extended constitutive relation~\cite{Ciraci16}
\begin{equation}
    \label{eq:theory_schdm2}
    \frac{e n_0(\mathbf{r})}{m_e} \nabla \left( \frac{\delta G[n]}{\delta n} \right)_1+ \omega(\omega + i \gamma)\mathbf{P}(\mathbf{r}) = -\varepsilon_0 \omega_p^2(\mathbf{r}) \mathbf{E}(\mathbf{r}),
\end{equation}
where $e$ is the elementary charge, $m_e$ is the electron mass, $\varepsilon_0$ is the vacuum permittivity, $\mathbf{r}$ is an observation point, $n_0(\mathbf{r})$ is the position-dependent ground electron density, $\gamma$ is the damping rate in the bulk material, $\mathbf{P}$ is the polarization density of the free electrons, $\omega_p(\mathbf{r})=\sqrt{e^2n_0(\mathbf{r})/(m_e \varepsilon_0)}$ is the position-dependent plasma frequency, and $\mathbf{E}$ is the electric field. $G$ is the internal energy functional, which takes the electron--electron interaction into account. The first term acts as a quantum correction to an otherwise very familiar classical relation~\cite{Griffiths13}. Depending on the level of sophistication of $G$, a hierarchy of semiclassical models can be built. For example, if $G$ is the Thomas-Fermi (TF) kinetic energy and $n_0(\mathbf{r})=n_0 \mathbf{1}_V(\mathbf{r})$, where $n_0$ is the bulk ground-state electron density and $\mathbf{1}_V$ is the indicator function, assuming the value $1$ as long as $\mathbf{r}$ is in the volume $V$ of the scatterer and $0$ otherwise, Equation~(\ref{eq:theory_schdm2}) yields the (hard-wall) HDM~\cite{Ciraci16}. The standard SC-HDM consists additionally of the von Weizs\"{a}cker (vW) gradient correction (in order to consistently account for spatial variations in the ground-state density) and an XC contribution, with both Local and Nonlocal Density Approximations (LDA and NLDA, respectively) having been presented~\cite{Toscano15,Ciraci16}. In particular~\cite{Ciraci16},
\begin{equation}
    \label{eq:theory_G}
    G[n] = T_{\rm TF}[n] + \frac{1}{\eta}T_{\rm vW}[n] + E_{\rm XC}[n],
\end{equation}
where for the XC term, we use the Perdew-Zunger LDA parametrization of the XC potential, following~\cite{Ciraci16}. The parameter $\eta$ is the weight of the vW term. In the limit $\eta \rightarrow \infty$, we recover the TF kinetic energy. Typically, $\eta \in[1,9]$~\cite{Khalid20}. Here, we adopt $\eta=1$, which is the well-suited for rapidly varying electron densities (or alternatively, for perturbations of short wavelength)~\cite{Ciraci16,Jones71}. Note that this is the value in the original derivation of von Weizs\"{a}cker~\cite{Weizsacker35}. In other works~\cite{Toscano15,Khalid21}, the value $\eta = 9$ is used, which is the asymptotic result for slowly varying electron densities~\cite{Ciraci16}. Alternatively, it is possible to extract $\eta$ as a best-match variable in comparison with results from DFT simulations~\cite{Yan15,Zhou23}.
\begin{figure}[h!]
    \centering
    \includegraphics[scale=0.9]{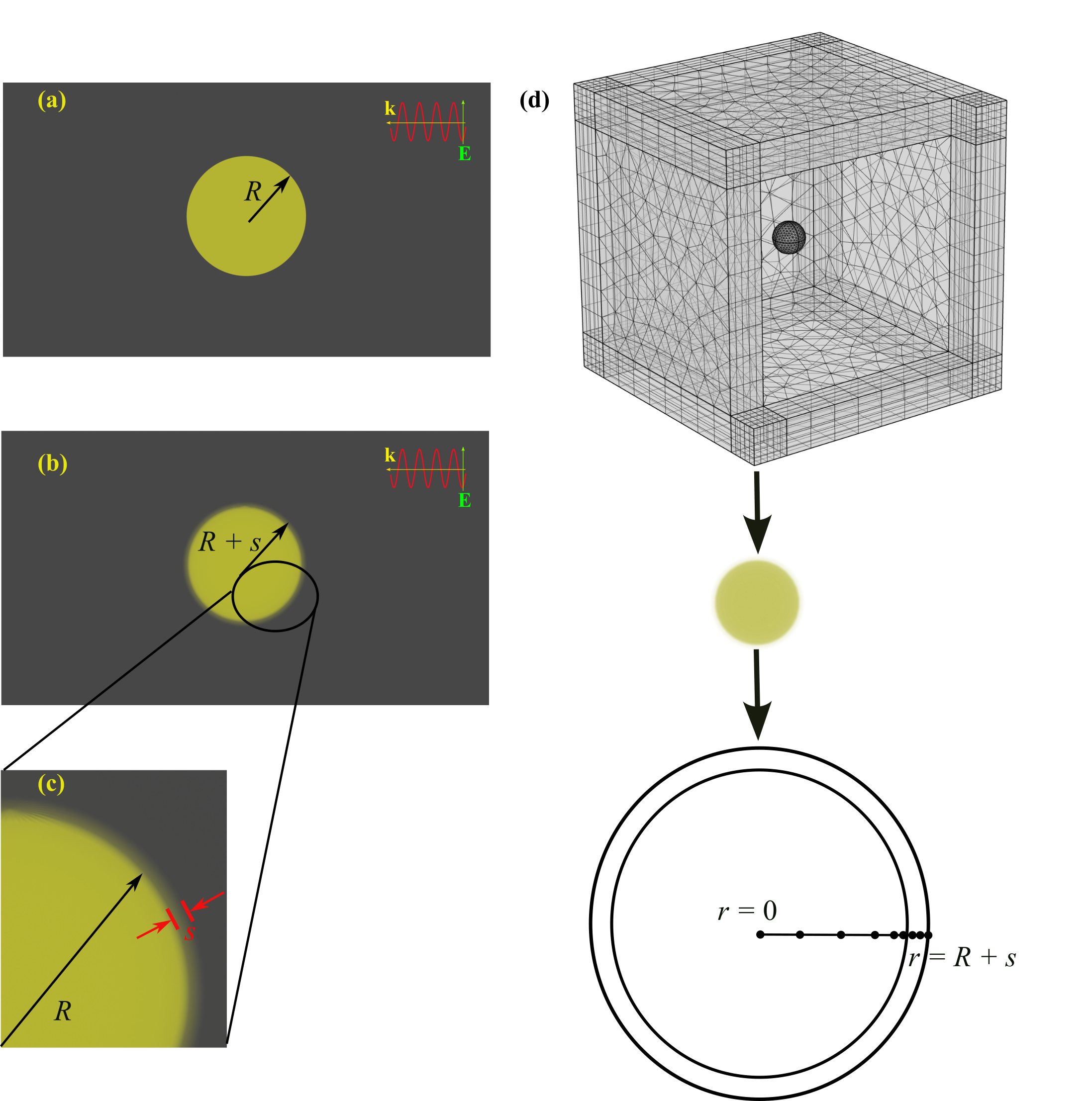}
    \caption{(a) A local spherical NP of radius $R$ is illuminated by a plane wave described by electric field $\mathbf{E}$, while the direction of propagation is denoted by the wave vector $\mathbf{k}$. Notice the well-defined boundary, due to the hard-wall assumption. (b) A spherical NP described by SC-HDM. (c) Zoom-in to depict better the spill-out. A soft shell wraps around the NP's hard ionic core, which is described by the jellium model and terminates at $r=R$. The shell, representing the decaying electron tail beyond the geometric boundary, formally extends to infinity, but practically is assigned a thickness $s$, the spill-out allowance. (d) Evolution of the treatment of the SC-HDM NP, from a DE-based perspective to an IE-based one and finally one exploiting the symmetries of the spherical geometry. In the first, careful meshes must be designed. The NP terminates at a distance $r=R+s$ from the origin. This nanosphere must be very finely discretized, in order to capture the wavelength of the matter waves of electrons (in the order of a few Angstr\"{o}m~\cite{Mortensen21}). Beyond the nanosphere extends a dielectric medium of zero electron ground-state and induced electron density. This medium is physically infinite; in DE-based methods a finite box is required. An adaptive mesh, dense in the near-field region (in order to capture its fast variations) and sparser far from it is typically used. To minimize the error of this abrupt truncation, Perfectly Matched Layers (PMLs) are employed, which must be also properly meshed. Our VIE does not enter into such discussions, discretizing only the nanosphere and encapsulating the infinite background in the Green's dyad. However, even this meshing redundant in light of the symmetries of the problem: The use of basis and testing functions such as the ones of Equations~(\ref{eq:theory_basis}) and (\ref{eq:theory_testing}) enables significant analytical progress and leads to the possibility of 1D discretization, as suggested in the bottom schematic (the two concentric circles represent the jellium edge and the truncation of the NP, from in to out.}
    \label{fig:theory}
\end{figure}

The expressions for the functionals and the functional derivatives of $G$ are quite complicated and available in the literature~\cite{Ciraci16}. We collect here compact expressions for the latter; the numerical treatment thereof depends exclusively on the order of derivative of $n_1(\mathbf{r}) = \frac{1}{e}\nabla\cdot\mathbf{P}(\mathbf{r})$---the induced electron density---they contain. In particular
\begin{equation}
    \label{eq:theory_Gfun}
    \left(  \frac{\delta G[n(\mathbf{r})]}{\delta n } \right)_1 = 
    \left[ \gamma_{\rm TF}(\mathbf{r} ) + \gamma_{\rm XC} (\mathbf{r}) +
    \gamma_{\rm vW,1}(\mathbf{r} )   \right] n_1( \mathbf{r}) + 
    \pmb{\gamma}_{\rm vW,2}(\mathbf{r}) \cdot \nabla n_1( \mathbf{r}) + 
    \gamma_{\rm vW,3}(\mathbf{r}) \nabla^2 n_1(\mathbf{r}),
\end{equation}
where by $\gamma$ we denote material kernels (we include more details in the Supporting Information--SI), whose exact form depends on the functional used in $G$---and by extension on $n_0(\mathbf{r})$. It is clear that the TF and XC (within the LDA) terms contribute exclusively functional derivatives that depend on $n_1(\mathbf{r})$; solution strategies for the HDM (TF term only) can be readily generalized for all contributions in this description level, and indeed such works have recently appeared~\cite{Du24}. The additional complexity is due to the vW term and translates to existence of higher-order derivatives of $n_1(\mathbf{r})$ in Equation~(\ref{eq:theory_Gfun}) (note in passing that extensions of the SC-HDM which include NLDA contributions to the XC term, as in~\cite{Toscano15}, can be treated in the same scheme).

From Equations~(\ref{eq:theory_schdm2})-- (\ref{eq:theory_Gfun}), it is clear that the major input material response function is the ground-state density $n_0(\mathbf{r})$, reflecting a departure from the hard-to-define in the nanoscopic regime bulk permittivities of classical electrodynamics. To simulate electron spill-out, we use a simple sigmoid function borrowed from~\cite{Mortensen21b}, namely
\begin{equation}
    \label{eq:theory_density}
    n_0( r ) = \frac{n_0}{2}\left[ 1- \tanh{\left( \frac{r-R}{a_n}\right)}\right].
\end{equation}
This model is particularly well-suited for spherical NPs, and we show such geometry (together with the excitation) in Figure~\ref{fig:theory}.
In Equation~(\ref{eq:theory_density}), $R$ is the radius of the spherical NP (here defined as the termination of the ionic lattice of the metal, where the ionic ground-state density falls to zero, following the jellium model), and $a_n$ is the decay parameter, determining how fast or slow the electron density decays across the interface. This parameter is critical to all calculations. Here, we match it to the model decay extracted in~\cite{Ciraci16} (logistic in that work), which in turn reproduces the decay of ground-state densities from full Kohn-Sham DFT calculations; see the SI for a detailed discussion on the ground-state density. The value is $a_n=2a_0/1.05$, where $a_0$ is the Bohr radius, which we use throughout. A model calculation of $n_0(\mathbf{r})$ is the most computationally efficient approach (and one that does not sacrifice accuracy~\cite{Ciraci16}). Alternatively, and to bypass DFT calculations, it is possible to extract $n_0(\mathbf{r})$ \emph{self-consistently}, that is, from the same equation of motion (taken at equilibrium) that Equation~(\ref{eq:theory_schdm2}) originates from~\cite{Toscano15}. This approach is called \emph{Orbital-Free} (OF) DFT~\cite{Mi23}.

A notable feature of SC-HDM is the effective abolition of boundaries. The reason is the introduction of electron spill-out. In that light, Equation~(\ref{eq:theory_schdm2}) applies to the \emph{whole domain}, since the NP extends (nominally) to infinity (see Figure~\ref{fig:theory}(a) and (b)), albeit with an exponentially decaying ground-state density profile~\cite{Ciraci16}. From a Computational Electromagnetics (CEM) perspective, this is a profound complication; solution strategies have been designed under the assumption that the (closed) scatterer is \emph{finite}, the jump of discontinuity of the material response functions translating to appropriate boundary conditions. In practice, and motivated by the exponential decay of $n_0(\mathbf{r})$, the NP extends up until a finite distance from its geometric edge, 
namely the thickness of the spill-out region, which we name \emph{spill-out allowance} $s$. From a computational perspective, this parameter is critical to the stability of simulations, as SC-HDM tends to be very sensitive with respect to it~\cite{Ciraci16,Baghramyan21} and our method confirms such dependence (see the SI). This is true beyond the so-called \emph{critical frequency}, which has been characterized as the frequency threshold of the photoelectric effect~\cite{Yan15}. Then, SC-HDM predicts induced charge densities that have a damped oscillatory behavior~\cite{Ciraci16,Yan15}, conflicting with the common (and reasonable) boundary condition which stipulates vanishing of the induced polarization density at the spherical interface $r=R+s$. We follow this boundary condition with one small caveat: we do not enforce such a behavior in the numerical polarization density, but remove from the formalism boundary terms. Thus, the polarization density follows the dynamics enforced by SC-HDM (see the SI for a discussion on the two approaches). Modified boundary conditions have been suggested~\cite{Yan15}, however the solution seems to be higher-order corrections to the kinetic energy functional~\cite{Baghramyan21}, which introduce in tandem significant complexity in the formalism. The sensitivity of SC-HDM on $s$ is a characteristic of the model and we design our approach around it rather than trying to resolve it (by introducing, e.g., size-dependent damping) in this work.

\subsection{Electrodynamics: Dyadic Green's Function Formalism}
\label{subsec:electro}

Irrespective of the sophistication of the semiclassical material description of Equation~(\ref{eq:theory_schdm2}), the electromagnetic (EM) wave dynamics is described by 
\begin{equation}
    \label{eq:theory_wave_e}
    \nabla \times \nabla \times \mathbf{E}(\mathbf{r}) - \frac{\omega^2}{c^2}\mathbf{E}(\mathbf{r}) = \omega^2 \mu_0 \mathbf{P}(\mathbf{r}),
\end{equation}
where $c$ is the speed of light in vacuum. We decompose the total field $\mathbf{E}(\mathbf{r})$ in terms of incident and scattered field components~\cite{Jackson75}
\begin{equation}
    \label{eq:theory_decomp}
    \mathbf{E}(\mathbf{r}) = \mathbf{E}_{\rm sca}(\mathbf{r}) + \mathbf{E}_{\rm inc}(\mathbf{r}),
\end{equation}
where $\mathbf{E}_{\rm sca}(\mathbf{r})$ satisfies a wave equation very similar to Equation~(\ref{eq:theory_wave_e}) and $\mathbf{E}_{\rm inc}(\mathbf{r})$ an equation very similar to its homogeneous version. Essentially, we delegate the generation of the scattered field to an induced current density (defined as $\mathbf{J}(\mathbf{r})=-i\omega\mathbf{P}(\mathbf{r})$), which occupies the same volume as the NP (including spill-out) rather than to the material contrasts due to the decaying $n_0(\mathbf{r})$. This is the well-known \emph{volume equivalence principle}~\cite{Chew95}, applied here to SC-HDM.

To formally relate the scattering field with the induced polarization density, we use the Green's function formalism, in particular~\cite{Chew95}
\begin{equation}
    \label{eq:theory_dyadic_int}
    \mathbf{E}_{\rm sca}(\mathbf{r}) =  \omega^2 \mu_0 \int_{V'} \overline{\mathbf{G}}(\mathbf{r},\mathbf{r}') \cdot \mathbf{P}(\mathbf{r'})d\mathbf{r}'.
\end{equation}
Above, $\mathbf{r}'$ denotes the position vector of sources and $V'$ is the volume which supports the polarization density, i.e., the volume of the NP. The problem of EM scattering is here interchanged by that of radiation from the polarization density at $V'$.  It is clear that the success of the method hinges upon the availability of the dyadic Green's function $\overline{\mathbf{G}}(\mathbf{r},\mathbf{r}')$ in closed form. For the case of spherical geometries it is available~\cite{Chew95},
\begin{equation}
    \label{eq:theory_dyad}
    \overline{\mathbf{G}}( \mathbf{r}, \mathbf{r}') = i k_0 \sum_{p=1}^{\infty} \sum_{q=-p}^{p} \frac{1}{p(p+1)} \left[ \mathbf{M}_{pq}( \mathbf{r}) \mathbf{M}_{pq}^*( \mathbf{r}') + \mathbf{N}_{pq}( \mathbf{r}) \mathbf{N}_{pq}^*( \mathbf{r}')  \right] - 
    \frac{\mathbf{\hat{r}\hat{r}}}{k_0^2} \delta(\mathbf{r} - \mathbf{r}'),
\end{equation}
where $k_0 = \omega/c$ is the vacuum wavenumber, $p$ and $q$ are the spherical summation indices, $\mathbf{M}_{pq}(\mathbf{r})$ and $\mathbf{N}_{pq}(\mathbf{r})$ are the vector wave functions~\cite{Stratton41}, $\delta$ is the Dirac delta function, and $\hat{\mathbf{r}}\hat{\mathbf{r}}$ is the dyadic product between the unit vectors of the radial direction. $\mathbf{M}_{pq}(\mathbf{r})$ and $\mathbf{N}_{pq}(\mathbf{r})$ contain spherical Bessel or Hankel functions of the first kind; which one appears above depends on the relative distance between observation and source point~\cite{Chew95}. With $*$ we denote the complex conjugate over $(\theta,\phi)$. The second term of Equation~(\ref{eq:theory_dyad}) reflects the well-known singularity of the Green's function when observation and source points overlap.

Using the decomposition of Equation~(\ref{eq:theory_decomp}) in Equation~(\ref{eq:theory_schdm2}) and subsequently employing Equation~(\ref{eq:theory_dyadic_int}) is the decisive step to graduate from the DE-based outlook of previous works~\cite{Ciraci16,Takeuci22,VidalCodina23} towards an IE-based formalism. The working equation of the proposed approach is
\begin{equation}
    \label{eq:theory_efie}
    -\frac{1}{e} \nabla
    \left( \frac{\delta G[n]}{\delta n} \right)_1 + \frac{1}{\varepsilon_0 \chi(\mathbf{r};\omega)} \mathbf{P}(\mathbf{r}) - \omega^2 \mu_0  \int_{V'} \overline{\mathbf{G}}( \mathbf{r}, \mathbf{r}') \cdot \mathbf{P}(\mathbf{r'})d\mathbf{r'} = \mathbf{E}_{ \rm inc} (\mathbf{r}),
\end{equation}
where $\chi(\mathbf{r},\omega)$ is the spatially dependent Drude susceptibility.
When $G[n]=0$, the case pertinent to LRA, we recover the well-known Volume Integral Equation (VIE) of classical electrodynamics~\cite{Jin15}. We do not transition from a DE-based to an IE-based approach out of mere curiosity. The proposed simulation approach, a VIE for SC-HDM, discretizes \emph{only} the volume of the NP (or volumes of aggregates thereof), which is the hotspot of quantum plasmonic activity (the infinite space being accounted for natively and exactly by the dyadic Green's function), instead of the entire domain (which requires truncations of the infinite space and appropriate absorbing conditions to minimize artifacts). We represent such shift in mentality in Figure~\ref{fig:theory}(c). Conceptually, it introduces a similar paradigm shift that Boundary Integral Equation (BIE) methods brought to the simple HDM, a decade ago~\cite{Yan13,Zheng18}. There, the very existence of hard boundaries allowed for a simpler approach, focusing exclusively on the surface of the nonlocal scatterers.

\subsection{Discretization Scheme: Symmetry-inspired Basis and Testing Functions}
\label{subsec:mie}

Integral equations are solved in classical CEM by means of a weighted residuals technique known as Method of Moments (MoM)~\cite{Harrington93}. The first step consists of expanding the sought after function, here $\mathbf{P}(\mathbf{r})$, in terms of an appropriate basis; the well-balanced choice between computationally simple and physically accurate local representations (basis functions) is more an art than an exact science~\cite{Newman91}. Here, inspired by the spherical symmetry of the NP, we resolve the angular dependence of $\mathbf{P}(\mathbf{r})$ in terms of \emph{vector spherical harmonics}, while delegating the radial dependence to Lagrange basis polynomials~\cite{NIST33}, leading to
\begin{equation}
    \label{eq:theory_expansions_b}
    \mathbf{P}(\mathbf{r}) = \sum_{i=1}^{N}\sum_{l=1}^{\infty} \sum_{m=-l}^{l} I_{i|l,m}^{(1)}\mathbf{g}_{i|l,m}^{(1)}(\mathbf{r})+I_{i|l,m}^{(2)}\mathbf{g}_{i|l,m}^{(2)}(\mathbf{r})+I_{i|l,m}^{(3)}\mathbf{g}_{i|l,m}^{(3)}(\mathbf{r}),
\end{equation}
where $i=1,\dots,N$ is the number of basis functions used, $l$ and $m$ are the usual spherical indices, $I_{i|l,m}^{(1)},I_{i|l,m}^{(2)},I_{i|l,m}^{(3)}$ are the weights of the expansion, which are to be calculated, and $\mathbf{g}_{i|l,m}^{(1)},\mathbf{g}_{i|l,m}^{(2)},\mathbf{g}_{i|l,m}^{(3)}$ are the basis functions, defined as
\begin{equation}
    \label{eq:theory_basis}
    \begin{cases}
        \mathbf{g}_{i|l,m}^{(1)}(\mathbf{r}) = N_{i}(r) \mathbf{Y}_{lm}(\theta,\phi), \\
        \mathbf{g}_{i|l,m}^{(2)}(\mathbf{r}) = N_{i}(r) \mathbf{Z}_{lm}(\theta,\phi), \\
        \mathbf{g}_{i|l,m}^{(3)}(\mathbf{r}) = N_{i}(r) \mathbf{X}_{lm}(\theta,\phi).
    \end{cases}
\end{equation}
In turn, $\mathbf{Y}_{lm}(\theta,\phi),\mathbf{Z}_{lm}(\theta,\phi),\mathbf{X}_{lm}(\theta,\phi)$ are the three vector spherical harmonics and $N_i(r)$ are the Lagrange basis polynomials. The derivatives applying on $\mathbf{P}(\mathbf{r})$, evident from Equations~(\ref{eq:theory_Gfun}) and (\ref{eq:theory_efie}), increase the smoothness requirements of the basis functions, and as such we opt here for cubic polynomials (see the SI for a detailed discussion of the basis functions and a slightly different indexing system than the one we use here).

To minimize the error of the truncation of the expansion in Equation~(\ref{eq:theory_expansions_b}), we project all terms of Equation~(\ref{eq:theory_efie}) on appropriate testing functions and require the result to be zero; the coefficients $I_{i|l,m}^{(1)},I_{i|l,m}^{(2)},I_{i|l,m}^{(3)}$ are calculated under this condition. The choice of testing functions is equally nontrivial. A common choice is to use the same functional format for both (the so-called Galerkin testing); in our case such a choice is promoted by the rich orthogonality properties that vector spherical harmonics satisfy~\cite{Feshbach53}. As such,
\begin{equation}
    \label{eq:theory_testing}
    \begin{cases}
        \mathbf{f}_{j|l',m'}^{(1)}(\mathbf{r}) = N_{j}(r) \mathbf{Y}_{l'm'}(\theta,\phi), \\
        \mathbf{f}_{j|l',m'}^{(2)}(\mathbf{r}) = N_{j}(r) \mathbf{Z}_{l'm'}(\theta,\phi), \\
        \mathbf{f}_{j|l',m'}^{(3)}(\mathbf{r}) = N_{j}(r) \mathbf{X}_{l'm'}(\theta,\phi),
    \end{cases}
\end{equation}
where $j=1,\dots,N$ is the number of testing functions used, and $l'$ and $m'$ denote the spherical indices. We underline that the particular choice of basis/testing functions is absolutely critical to the simplification of our method. The aforementioned orthogonality properties allow for significant analytical progress and essentially eliminate the need to numerically perform multidimensional integrals. We perform numerical integrations only with respect to the $r$ variable; the problem and the discretization thereof become effectively \emph{one-dimensional}. This final step of our simplified modeling is shown in Figure~\ref{fig:theory}(c). By this token, the discretization of a linear segment extending from $r=0$ to $r=R+s$ is trivial, though we opt for an inhomogeneous sampling (see Figure \ref{fig:theory}(d)) in order to acknowledge the spatial variation of $n_0(\mathbf{r})$ (with sparser discretization in the bulk and denser in the selvage region, where $n_0(\mathbf{r})$ varies rapidly).

The expansion of $\mathbf{P}(\mathbf{r})$ in a set of appropriate basis functions, its projection on testing functions, and subsequent numerical integration, transforms Equation~(\ref{eq:theory_efie}) into a system of linear equations as follows 
\begin{equation}
    \label{eq:theory_system}
    (\mathbf{TF} + \mathbf{XC} + \mathbf{vW}_1 + \mathbf{ML} + \mathbf{S}) \mathbf{I} - [(\mathbf{vW}_2 + \mathbf{vW}_3)\mathbf{AUX}]\mathbf{I} = \mathbf{b}.
\end{equation}
Equation~(\ref{eq:theory_system}) reflects the material contributions from Equation~(\ref{eq:theory_Gfun}). The terms $\mathbf{TF}$, $\mathbf{XC}$, and  $\mathbf{vW}$ arise due to the particular energy functional in Equation~(\ref{eq:theory_Gfun}). The vW term gives three separate contributions reflecting its dependence on $n_1(\mathbf{r})$, $\nabla n_1(\mathbf{r})$, and $\nabla^2 n_1(\mathbf{r})$. $\mathbf{ML}$ is the matrix due to the second term of Equation~(\ref{eq:theory_efie}) (we name it ``local'', since it appears also in the simple LRA) and $\mathbf{S}$ is the matrix due to the third term of Equation~(\ref{eq:theory_efie}) (which is the scattered-field term). $\mathbf{b}$ denotes the excitation; in this work, plane waves probe the spherical NPs (see Figures~\ref{fig:theory}(a) and (b)). However other excitations (e.g., spherical waves, point dipoles, electron beams) can be implemented. The matrix $\mathbf{AUX}$ reflects the introduction of an auxiliary variable in order to reduce the order of the derivatives applying on $n_1(\mathbf{r})$ (and as such expand the selection space of basis/testing functions; for more details, see the SI). Finally, $\mathbf{I}$ is the vector of the unknown coefficients in Equation~(\ref{eq:theory_expansions_b}); since we use three sets of basis functions, it is a $3N\times1$ vector, while all the system matrices are $3N\times3N$.

Equation~(\ref{eq:theory_system}) is the most essential result of this work. There are several aspects to discuss here. 
\begin{enumerate}
    \item Equation~(\ref{eq:theory_system}) does not merely describe a solution strategy for SC-HDM, but is rather a \emph{solver generator}, in the same vein that SC-HDM is a hydrodynamic-model generator, depending on the terms introduced in $G$. As long as $G=G[n,\nabla n]$ holds, the method detailed here can effectively discretize any material model (beyond the one we demonstrate herein). We exploit extensively this property of Equation~(\ref{eq:theory_system}), to apply our recipe to the LRA and the simple HDM in order to verify it in a \emph{modular manner} in Section~\ref{sec:res}.
    \item It can be shown that the coefficients $I_{i|l,m}^{(1)}$ and $I_{i|l,m}^{(2)}$ decouple from the coefficients $I_{i|l,m}^{(3)}$. Equation~(\ref{eq:theory_system}) can be simplified to two linear systems (see the SI), thus reducing the dimensions of the system matrix to (at maximum) $2N\times2N$. This simplification corresponds to the orthogonality between transverse electric (TE) modes from the one side and transverse magnetic (TM) and longitudinal modes from the other side (note that $I_{i|l,m}^{(1)}$ and $I_{i|l,m}^{(2)}$ correspond to $\mathbf{Y}_{lm}$ and $\mathbf{Z}_{lm}$, respectively, which in turn are  components of the spherical wave functions $\mathbf{N}_{lm}$ and $\mathbf{L}_{lm}$; on the other hand, $I_{i|l,m}^{(3)}$ corresponds to $\mathbf{X}_{lm}$ which in turn is component of the spherical wave functions $\mathbf{M}_{lm}$). This decoupling is well-known in LRA and HDM approaches~\cite{Mystilidis23a} and we now demonstrate it to SC-HDM as well.
    \item It is interesting that, even though the SC-HDM results in a far more complicated system than LRA, the most challenging computations are still monopolized by the classical terms. The scattering term ($\mathbf{S}$ in matrix notation) is dense and involves computationally demanding self-terms (organized along the diagonal of the matrix). Together with the local and the excitation term, they are the only frequency-dependent terms of Equation~(\ref{eq:theory_system}). On the contrary, the material terms that arise due to the introduction of quantum behavior end up sparse (see the SI) and independent of frequency: only a single calculation is necessary! From this perspective, we expect that SC-HDM \emph{should, in principle,} have similar (albeit increased) computational requirements to that of full LRA simulations. This is a reassuring statement to the early pessimism of the efficiency of the method~\cite{Toscano15,Stamatopoulou22} and a clear computational breakthrough concerning sophisticated (other than the HDM) hydrodynamic models.
\end{enumerate}

\subsection{Feibelman $d$ Parameters}
\label{subsec:obs}
Using the method detailed above, we numerically determine efficiently and accurately $\mathbf{P}(\mathbf{r})$. In principle, we can extract, in a straightforward manner, near-field distributions, cross sections (scattering, absorption, and extinction), and induced charge densities (and indeed we will demonstrate such calculations shortly). Here we will elaborate on the extraction of a quantity, namely Feibelman parameters, which is not directly observable, but facilitate further theoretical and numerical explorations.

Feibelman parameters are mesoscopic material response functions~\cite{Feibelman82} that capture the centroid of the induced charge density and the normal derivative of the induced current density tangential to the interface, denoted as $d_\perp$ and $d_\|$~\cite{Christensen17}. Typically, $|d_\perp| \gg |d_\||$~\cite{Yang19} (with $d_\|$ vanishing at charge-neutral jellium interfaces~\cite{Apell81}), and as such we discuss here only the extraction of the former (which is also the common strategy~\cite{Mortensen21b}). In the SRM, such functions are used to amend the conventional boundary conditions of electromagnetics (continuity of tangential components of the electric field, continuity of tangential components of the magnetic field). In this manner, nonlocality, surface-enhanced Landau damping, and electron spill-out are all captured on equal ground, in a computationally efficient mesoscopic manner. The materials of the scatterers however can still be modeled by means of conventional bulk permittivities (following the Drude or the Lorentz-Drude model), which constitutes  a major advantage. Nonetheless, the calculation of Feibelman parameters passes through detailed microscopic calculations (typically carried out for planar interfaces)~\cite{Liebsch87,Gonçalves20}, setting an effective computational bottleneck in all methods that address SRM. As we will demonstrate later, our VIE formalism of SC-HDM can be standalone very efficient for spherical NP, however, it can also be used to calculate Feibelman parameters as well, and then be combined with SRM methods that address more complicated geometries (e.g., aggregates of spherical NPs). The Feibelman parameter $d_\perp$ can be easily calculated by the induced electron density as follows~\cite{Baghramyan21}   
\begin{equation}
    \label{eq:theory_feib}
    d_\perp = \frac{\displaystyle\int_0^{R+s} n_1(r) (r-R) r^2  dr}{\displaystyle\int_0^{R+s} n_1(r) r^2  dr}.
\end{equation}

\section{Results}
\label{sec:res}

\begin{figure}[t!]
    \centering
    \includegraphics[width=0.9\linewidth]{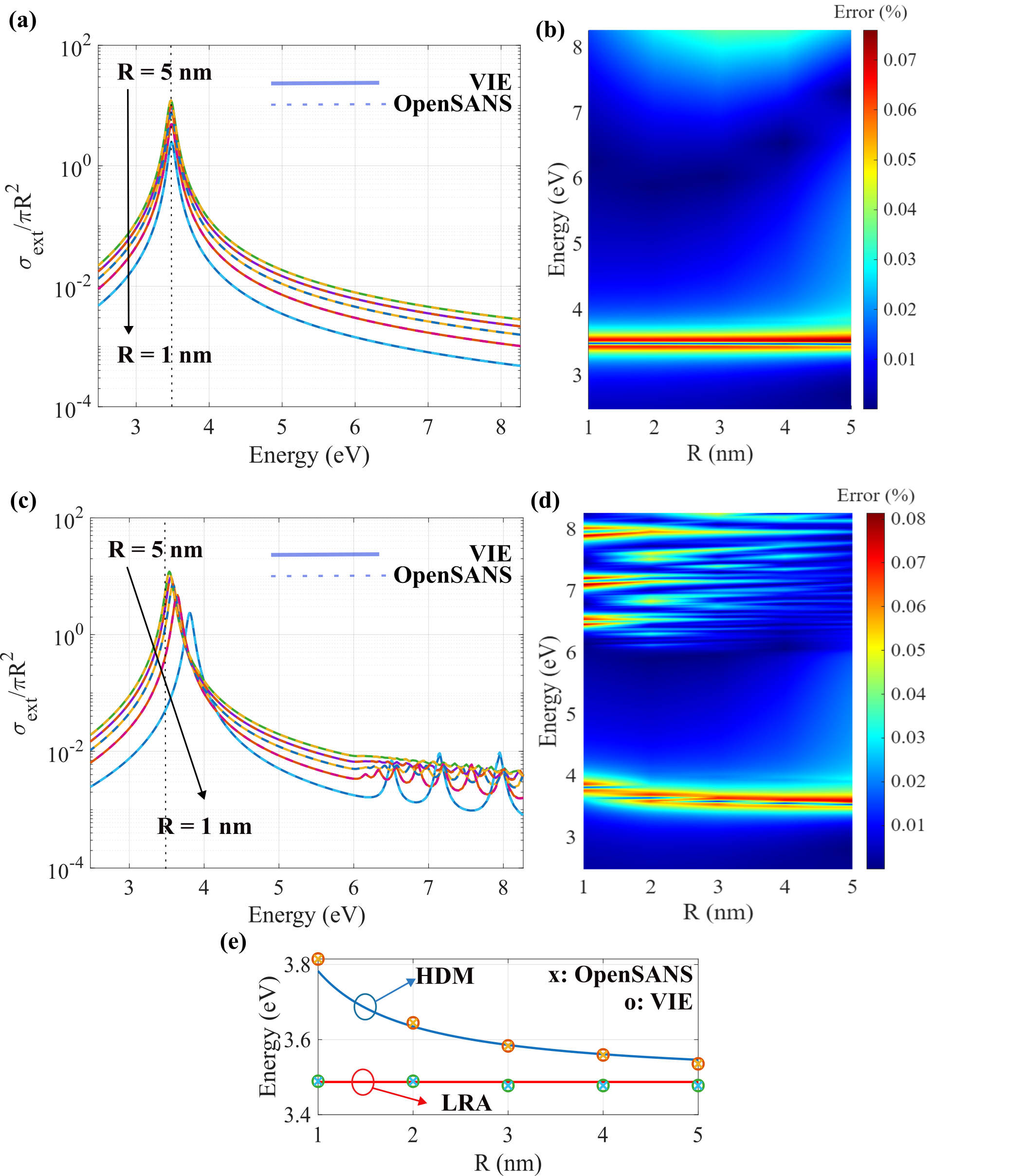}
    \caption{Validation of our computational scheme through OpenSANS and quasistatic theory.  (a) Extinction spectra of Na spherical NPs, normalized to the geometrical cross-section $\pi R^2$, within the LRA description. The dashed line indicates the LRA quasistatic result $\omega_p/\sqrt{3}$. (b) Relative error between the extinction spectra of the VIE approach and OpenSANS regarding LRA. (c) Extinction spectra of Na spherical NPs, normalized to the geometrical cross-section $\pi R^2$, using the HDM description. The dashed line indicates the LRA quasistatic result $\omega_p/\sqrt{3}$. (d) Relative error between the extinction spectra of the VIE approach and OpenSANS regarding HDM. (e) Comparison of the dipolar LSP as a function of the NP radius between OpenSANS (crosses), the VIE approach (circles) for the LRA and the HDM with the relevant quasistatic results (LRA is represented by the red line and HDM by the blue one).}
    \label{fig:res_lra_hdm}
\end{figure}

The method proposed here serves as a solution technique not only for SC-HDM, but also for less sophisticated approaches, such as the classical LRA and the hard-wall HDM. As such, it is possible to validate it against alternative methods, pertinent to such models. Here we use as a reliable method to compare with OpenSANS~\cite{Mystilidis23a}. OpenSANS can treat multilayered planar, circular cylindrical, and spherical structures, relying on a scattering-matrix formalism~\cite{Benedicto15,Dong16,Dong17}. This is a semi-analytical method, which is very efficient, highly accurate and we consider it an excellent benchmark platform for our numerical methods (see also~\cite{Mystilidis23b,Zheng22}).

We collect our main results in Figure~\ref{fig:res_lra_hdm}. We simulate five Na nanospheres with decreasing radii ($5$, $4$, $3$, $2$, and $1$ nm). The fabrication of noble-metal nanospheres of such dimensions is possible, and the spherical symmetry is maintained even for the smallest among them~\cite{Scholl12}; in terms of the material composition, Na is notoriously reactive~\cite{Lindau71} and tough to fabricate. However it represents a very realistic ``free--electron metal'' (for which SC-HDM was designed). The NPs are probed by a plane wave (as in Figure~\ref{fig:theory}), and we collect the normalized (to the geometric) extinction cross section (the full simulation setup can be found in the SI).
In Figure~\ref{fig:res_lra_hdm}(a), we compare using LRA as the material model, capturing in both methods the expected dipolar localized surface plasmon (LSP) resonance, at about $3.49$\,eV, in agreement with quasistatic theory. The agreement between the two approaches is excellent; the relative error is shown in Figure~\ref{fig:res_lra_hdm}(b). We notice that there is no size dependence with respect to the dipolar LSP frequency, only amplitude diminution with decreasing radius, as expected~\cite{Bohren98}.

In Figure~\ref{fig:res_lra_hdm}(c), we compare the two solvers using the HDM as the material model, and show the relative error in Figure~\ref{fig:res_lra_hdm}(d). The agreement is good again, and we notice here the trademark characteristics of nonlocal optical response, namely a blueshift of the dipolar LSP as the dimensions decrease~\cite{Raza13a} (compare with the dashed line which corresponds to the quasistatic LRA result) and, beyond the plasma frequency (here $6.04$ eV, see also the SI), the manifestation of additional resonances, originating from the propagation of additional longitudinal degrees of freedom in the material~\cite{Raza11}. It is important to note that both solvers agree even in these extremely high frequencies, where the mesh requirement for our VIE becomes quite restrictive. Nonetheless, due to the 1D nature of our modeling, we can circumvent the problem easily and capture the longitudinal resonances accurately with very few elements (linear segments, see the SI for more details on the mesh).

We finally collect the position of the dipolar LSP resonance in both the LRA and HDM cases, using our VIE approach and OpenSANS and contrast them with the asymptotic-analytical result from quasistatic theory in Figure \ref{fig:res_lra_hdm}(e). In the classical theory, the nanosphere can be approximated well by a point dipole~\cite{Bohren98} oscillating at $\omega_p/\sqrt{3}$. On the other hand, when spatial dispersion is admitted through the HDM, the dipole resonance manifests a $1/R$ dependence~\cite{Raza15}, in particular a blueshift,
\begin{equation}
    \label{eq:res_omega_hdm}
    \omega_{\rm LSP} \sim \frac{\omega_p}{\sqrt{3}} + \frac{\sqrt{2} \beta}{2R},
\end{equation}
where $\beta$ is the strength of the hydrodynamic correction. VIE approach and OpenSANS agree excellently with each other and very well with the asymptotic results in both material models; we underline that the evidently stronger blueshift for small radii of both solvers in HDM and in comparison with the theory comes from the fact that they provide fully retarded results~\cite{Raza15}.

The agreement between the independent methods and the quasistatic approach, and the prediction of all the relevant physics in these two setups, are strong validations of our method. Recalling the modular manner the functional $G$ and by extension our linear system of Equation~(\ref{eq:theory_system}) are formulated, the validation performed above is essentially matrix-by-matrix: The LRA setup confirms that the implementation of matrices $\mathbf{ML}$ and $\mathbf{S}$ (and $\mathbf{b}$) are correct, while the HDM setup further validates the $\mathbf{TF}$ term.

\begin{figure}[h!]
    \centering
    \includegraphics[width=\linewidth]{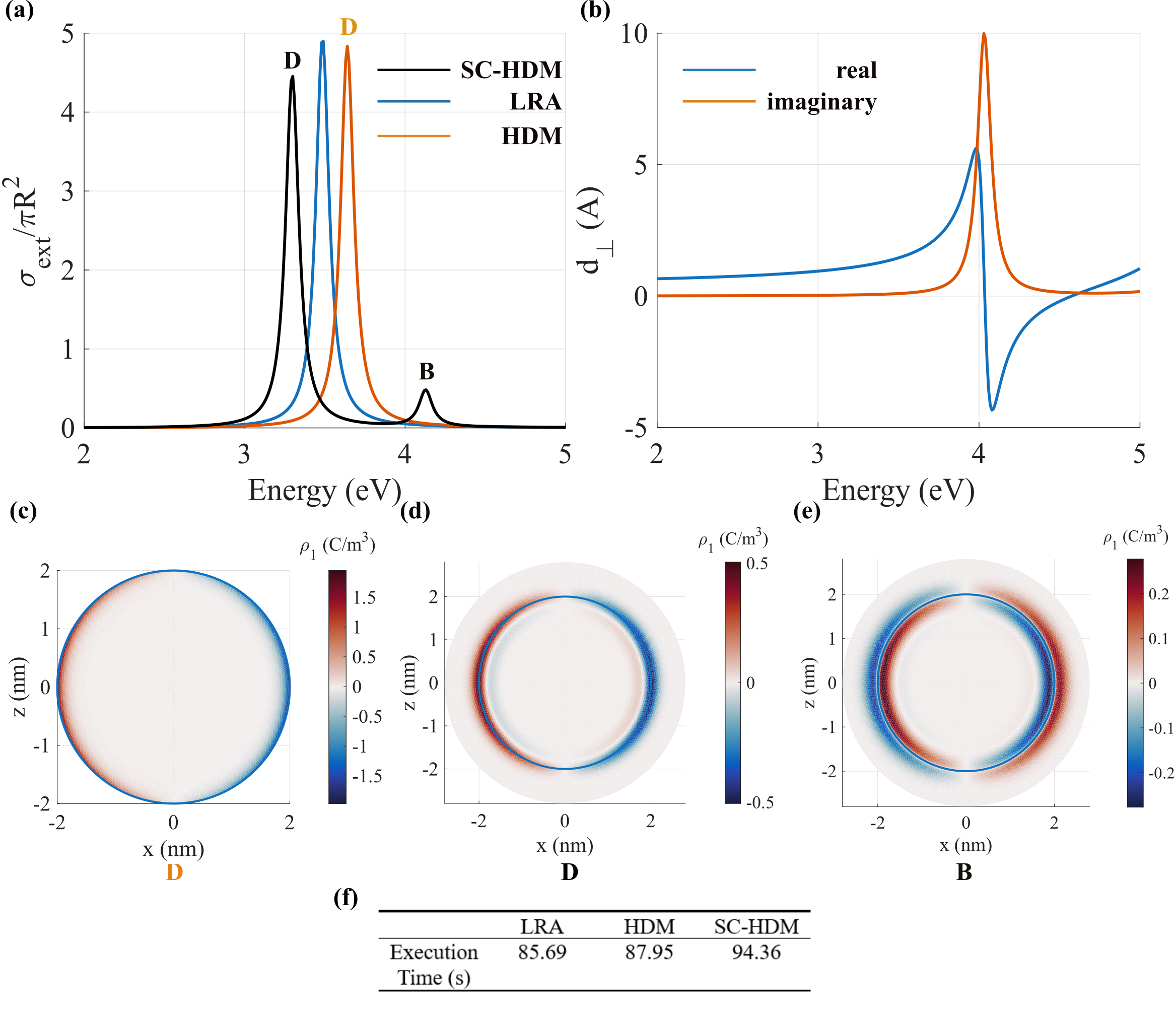}
    \caption{Comparison of results between LRA, HDM, and SC-HDM. (a) Normalized extinction cross sections within LRA (blue line), HDM (orange line), and SC-HDM (black line) Dipolar LSPs are marked with a D, while the Bennett resonance is marked with a B. (b) The Feibelman parameter $d_\perp$ retrieved from the SC-HDM calculation. (c) Induced charge density for the dipolar LSP resonance of HDM (orange D in panel (a)). (d) Induced charge density for the dipolar LSP resonance of SC-HDM (black D in panel (a)). (e) Induced charge density for the Bennett resonance of SC-HDM (black B in panel (a)). The blue circle is a guide to the eye which corresponds to the end of the nanosphere for HDM and the jellium edge for SC-HDM. (f) Simulation time for each model (averaged over $20$ runs).}
    \label{fig:comparison}
\end{figure}
With confidence in our approach, we study the archetypical Na nanosphere using all three material models: The LRA, the HDM, and the SC-HDM. We study a 2 nm nanosphere (see SI for full simulation details) and collect our results in Figure~\ref{fig:comparison}. We show the extinction cross sections (under a plane-wave excitation as in Figure~\ref{fig:theory}) for the three different models, normalized over the geometrical cross section. LRA shows a single distinct feature, the well-known dipolar LSP resonance. As before, the location thereof ($3.49$\,eV) agrees well with the classical quasistatic result $3.487$\,eV. The HDM shows a single feature as well (note that we probe here frequencies below the plasma one), the blueshifted dipolar LSP resonance (at $3.64$\,eV) in agreement with the literature~\cite{Raza13a,Raza15} and the extended quasistatic result, which yields $3.635$\,eV~\cite{Raza15}. Though correct in the framework of HDM, where the inclusion of the Fermi pressure augments the restoring force due to Coulomb interactions of LRA (and hence blueshifts the resonant frequency), the result is incorrect for Na nanospheres, as already discussed~\cite{Stella13}. There, the relatively low work function of the material leads to a pronounced electron spill-out~\cite{Toscano15}, which is neglected by the HDM. It is clear from Figure~\ref{fig:comparison}(a) that SC-HDM restores the correct result, that is, predicts a \emph{redshift} ($3.3$\,eV) of the dipolar LSP of the nanosphere, in accordance with previous work in SC-HDM~\cite{Toscano15,VidalCodina23}, theoretical~\cite{Teperik13a,Stella13,Liebsch93}, and experimental results~\cite{Brechignac92} for alkali metals. Most interestingly, a new feature beyond the dipolar LSP frequency (but below the plasma one) appears. This is the \emph{Bennett resonance}, which results due to the continuous decay of the electron density across the interface (see Equation~(\ref{eq:theory_density}))~\cite{Bennett70}. Though it is an achievement of SC-HDM that the Bennett resonance can be captured (and a validation to our solution strategy), we stress that it exists (with these material parameters) in a part of the spectrum where numerical computation is notoriously difficult: Previous FEM simulations~\cite{Ciraci16,Baghramyan21}, as well as our VIE method (see the SI for a detailed discussion) show significant dependence on the \emph{computational domain} and particularly on the spill-out allowance $s$ (a most bizarre finding for an IE-based method from a classical perspective!). This result, indigenous to SC-HDM rather than the computational method used to simulate it, calls for caution on the evaluation of results in this part of the spectrum (on the other hand, the dipole frequency is accurately captured). Works towards stabilizing this behavior have proposed going beyond conventional SC-HDM, that is, adopting more sophisticated $G$ functionals~\cite{Baghramyan21}, introducing size-dependent damping~\cite{Hu22} in an ad-hoc manner or in $G$ itself~\cite{Ciraci17}, or modifying the boundary conditions~\cite{Yan15}, though a universal solution has not been accepted yet. Our method, presented here for the most popular and fundamental prescription of SC-HDM, can be adapted to analyze such corrections in a more~\cite{Baghramyan21} or less~\cite{Ciraci17,Yan15,Hu22} laborious manner.

To clarify the nature of the resonances discussed above, we calculate the induced charge density at the (energy) location of the resonances and on a slice of the $xz$ plane, extending to $R$ (for HDM) or to $R+s$ (for SC-HDM). In Figure~\ref{fig:comparison}(c), we recognize the HDM dipole, with the charge density being pushed inside the hard boundary (electron spill-in), while in Figure~\ref{fig:comparison}(d) the SC-HDM dipole is shown. There, electron spill-out is very notable. Across the interface, the charge maintains its sign; the dipolar nature manifests on the NP \emph{as a whole}. For the Bennett resonance on the other hand, seen in Figure~\ref{fig:comparison}(e), the dipolar nature is seen \emph{across the interface}, which is the fingerprint of this mode.

Based on the induced charge density, we calculate finally the perpendicular Feibelman parameter $d_\perp$ (see Subsection~\ref{subsec:obs}), which captures the centroid of the induced charge density and is presented in Figure~\ref{fig:comparison}(b). It is clear that the dipole resonance does not cause structuring to $d_\perp$~\cite{Baghramyan21}. However, around the Bennett resonance a very notable structure is clear, that of a standard Lorentzian resonance. Essentially, the charge moves from outside the jellium edge to gradually inside due to Coulomb attraction. The second crossing of zero in Figure~\ref{fig:comparison}(b) at about $4.62$ eV  is unphysical and betrays the numerical issues we mentioned earlier: ${\rm{Re}}\{d_\perp\}$ rises to describe \emph{another Bennett resonance} situated at frequencies higher than the ones probed here. The proliferation of unphysical Bennett resonances is directly linked to SC-HDM's numerical issues (see the SI) and has been observed before~\cite{Baghramyan21}. We maintain that the calculation of such mesoscopic material response functions in the time scale of our method (see the table in Figure~\ref{fig:comparison}(f)) is very meaningful, as it prescribes a (definitely in its first steps) way out of computationally expensive TD-DFT calculations, which are currently needed to determine Feibelman parameters, that will then be fed to another semiclassical model, namely SRM. SC-HDM through our VIE method and once having reached its maturity, can offer meaningful Feibelman parameters to SRM solvers (which benefit from a structure much closer to that of LRA ones than a generic SC-HDM solver) in a matter of minutes.

We close this discussion with some additional remarks on the execution times for the three models we interrogated in Figure~\ref{fig:comparison}. $20$ runs under the same load where performed, timed from loading the inputs to the calculation of the $I$ coefficients (see Equation~(\ref{eq:theory_expansions_b})), the final results averaged to yield the table of Figure~\ref{fig:comparison}(f). It is clear that the performance is excellent. LRA and HDM run in very similar times. SC-HDM is undeniably computationally more expensive than the previous models. However the IE modeling presented here, together with the symmetry-selected basis functions, achieve execution times \emph{of the same order as in full classical simulations}, ran on a personal laptop for a quite sophisticated material response. This is, we believe, a significant contribution of this work. The efficiency just showcased reveals another use for our VIE for spherical NPs, that of a valuable benchmark platform for recipes addressing more complicated geometries. 

\section{Conclusion}

In this work, we present an innovative and efficient computational method for the advanced SC-HDM of mesoscopic plasmonics, which includes both nonlocality and electron spill-out in the material response, and suggest a significant shift in perspective, departing from DE-based models to an IE-based one. A VIE method is suggested, well suited for the scattering problems that monopolize the field and for the inhomogeneity of the SC-HDM NPs. For the case of spherical NPs, inspired by the highly symmetric nature thereof, we construct a system of basis and testing functions using extensively vector spherical harmonics. Using a MoM to solve the IE system with such basis/testing functions leads to an effectively 1D problem, as significant analytical progress becomes available. We validate our approach step-by-step, first for the classical LRA and then for the hard-wall HDM, using the semi-analytical software OpenSANS as well as theoretical results. We then perform a comparative study between the LRA, the HDM, and the SC-HDM. In this example, we reveal that our solution strategy captures all the distinctive physics of the models involved (blueshift with respect to LRA and spill-in for HDM, redshift with respect to LRA, Bennett resonance, and spill-out for SC-HDM). Notably this is achieved in virtually the same order of computational time irrespective of the complexity of the material response, highlighting the importance of our tool as a validation platform for alternative recipes pertinent to SC-HDM. In light of the performance of our method, we suggest that SC-HDM, through its ability to extract Feibelman parameters and after its computational instabilities are eliminated, may be integrated in a pipeline scheme with SRM solvers that are simpler to design, effectively substituting lengthy TD-DFT calculations.

\medskip
\textbf{Supporting Information} \par 
Supporting Information is available from the Wiley Online Library or from the author.

\medskip
\textbf{Acknowledgements} \par 
C. M. acknowledges support from the IEEE Antennas and Propagation Society Doctoral Research Grant and the  Research Foundation of Flanders (FWO), grant ID: V420424N. X. Z. acknowledges support from the IEEE Antennas and Propagation Society Postdoctoral Fellowship and from FWO, grant id: V408823N. C. M., \mbox{G. A. E. V.}, and X. Z. acknowledge support from KU Leuven (C1 project, id: C14/19/083; IDN project, id: IDN/20/014; small infrastructure grant, id: KA/20/019). The Center for Polariton-driven Light-Matter Interactions (POLIMA) is funded by the Danish National Research Foundation (Project No.~DNRF165).

\medskip

%
\bibliographystyle{MSP}
\bibliography{refs}


\begin{figure}
\textbf{Table of Contents}\\
\medskip
  \includegraphics{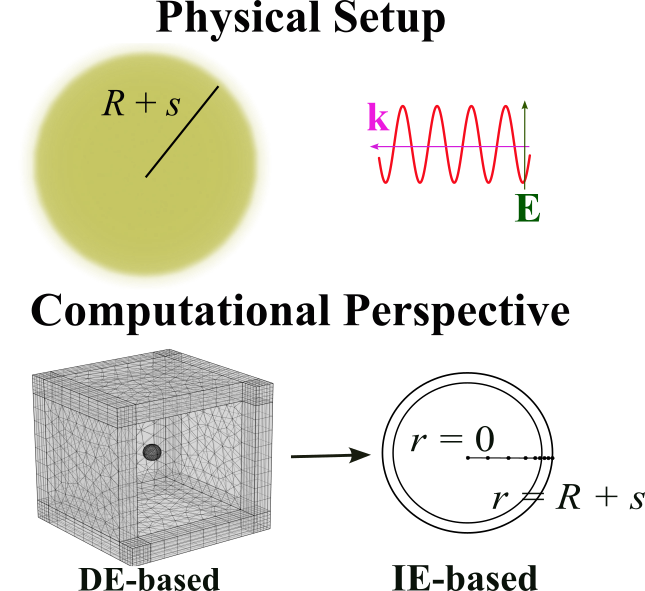}
  \medskip
  \caption*{Numerical modeling of mesoscopic material response models that capture the quantum dynamics of electrons does not have to come with discouraging computational bottlenecks. The main message is that through a shift in perspective in modeling towards Integral Equation methods and exploiting symmetry-based arguments, it is possible to capture complicated phenomena such as nonlocality and electron spill-out using a personal laptop.}
\end{figure}

\end{document}